\documentclass[usenatbib,useAMS]{mn2e}

\usepackage{epsf}
\usepackage[dvips]{epsfig}
\usepackage{times}
\newcommand{\oversim}[2]{\protect{\mbox{\lower0.5ex\vbox{%
  \baselineskip=0pt\lineskip=0.2ex
  \ialign{$\mathsurround=0pt #1\hfil##\hfil$\crcr#2\crcr\sim\crcr}}}}}
\newcommand{\simless} {\mbox{$\,\mathrel{\mathpalette\oversim<}\,$}} 

\begin{document}

\label{firstpage}

\title[Trapped disc stars]{Complex stellar populations in massive
  clusters: trapping stars of a dwarf disc galaxy in a newborn stellar
  super-cluster}

\author[Fellhauer et al.]{M. Fellhauer$^{{1,3}}$
  \thanks{madf@ast.cam.ac.uk}, P. Kroupa$^{2,3}$
  \thanks{pavel@astro.uni-bonn.de} and N. W. Evans$^{1}$
  \thanks{nwe@ast.cam.ac.uk} \\
  $^{1}$ Institute of Astronomy, University of Cambridge, Madingley Road, 
  Cambridge CB3 0HA, UK\\ 
  $^{2}$ Argelander Institute for Astronomy, University of Bonn, Auf dem
  H\"{u}gel 71, D-53121 Bonn, Germany\\
  $^{3}$ The Rhine Stellar-Dynamical Network} 

\pagerange{\pageref{firstpage}--\pageref{lastpage}} \pubyear{2006}

\maketitle

\begin{abstract}
  Some of the most massive globular clusters of our Milky Way, such as
  for example $\omega$-Centauri, show a mixture of stellar populations
  spanning a few~Gyr in age and $1.5$~dex in metallicities.  In
  contrast, standard formation scenarios predict that globular and
  open clusters form in one single star-burst event of duration
  $\simless 10$~Myr and therefore should exhibit only one age and one
  metallicity in its stars. Here, we investigate the possibility that
  a massive stellar super-cluster may trap older galactic field stars
  during its formation process that are later detectable in the
  cluster as an apparent population of stars with a very different age
  and metallicity.

  With a set of numerical N-body simulations, we are able to show
  that, if the mass of the stellar super-cluster is high enough and
  the stellar velocity dispersion in the cluster is comparable to the
  dispersion of the surrounding disc stars in the host galaxy, then up
  to about 40 per cent of its initial mass can be additionally gained
  from trapped disc stars. We also suggest that a super-cluster may
  capture in excess of its own mass under certain conditions.
\end{abstract}


\section{Introduction}
\label{sec:intro}

The most massive globular cluster in the Milky Way, $\omega$-Centauri,
\citep[$5 \times 10^{6}$~M$_{\odot}$,][]{mey95} is also one of the
most enigmatic of star clusters.  It shows a spread in stellar
populations in age ($\approx 3$--$5$~Gyr) and metallicity ($-2.0
\simless [\rm Fe/H] \simless -0.5$) \citep[e.g.][]{hil00,nor95} with
metal-poorer stars having a less-concentrated radial distribution than
the more metal-rich population \citep{leu00}. The metal-poorer
population also rotates with a maximum rotation speed of
$8$~km\,s$^{-1}$ \citep{fre01}, whereas the more metal-rich population
does not \citep[e.g.][]{nor97}. Seventy~per cent of the population of
$\omega$~Cen have a metallicity of [Fe/H]$\approx-1.7$, 25~per cent
have [Fe/H]$\approx-1.2$ while 5~per cent have [Fe/H]$\approx -0.7$
\citep{hil00}.  The cluster\footnote{The relaxation time of
$\omega$~Cen is close to a Hubble time and so it may be equally valid
to refer to this object as being a low-mass ultra-compact dwarf galaxy
(UCD).}  orbits the Milky Way in a slightly inclined (mostly within
the disc), highly eccentric (perigalacticon $\approx 1$~kpc,
apogalacticon $\approx 6.4$~kpc), retrograde orbit well within the
Solar radius \citep{tsu04}.  Some effects of the colour-magnitude
diagram (CMD) may only be explainable with an unusual helium
enrichment \citep{bek06,pio05} of some stars, while other workers
prefer the scenario of different distinct metallicity populations
\citep{sol06}.

A similar spread in metallicities is found in the most massive
globular cluster G1 in M31 \citep{mey01}.  It is even debated if
NGC~6388 and NGC~6441, two globular clusters of the Milky Way which
are unusually metal rich and quite massive, also show multiple stellar
populations \citep{ree02}. 

As a possible formation scenario, some authors believe that these
clusters are massive enough to retain part of their gas content,
stemming from their initial formation, which then is able to cause a
second episode of star formation \citep[e.g.][]{pla03}.  Another
scenario sees these clusters as stripped cores of disrupted nucleated
dwarf galaxies \citep[e.g.][]{bek06,ide04,tsu04}.  According to a
third scenario, a massive cluster may accrete co-moving inter-stellar
gas from its host galaxy leading to later episodes of star-formation
\citep{kro98}.  A fourth scenario proposed by \citet{kro98} sees a
massive star-cluster complex (a 'stellar super-cluster') forming in a
tidally-driven star-burst off-centre in the host galaxy thereby
capturing older and metal-poorer field stars from the host.  A simple
estimate shows that the captured population may be substantial if the
super-cluster formation-time-scale is shorter than the field-star
crossing time-scale through the forming super-cluster. \citet{fel04}
returned to this scenario in the context of $\omega$~Cen suggesting it
may have formed through merged star clusters from a star-cluster
complex during the star-burst in a disrupting gas-rich dwarf galaxy as
it plunged into the Milky Way.  During its formation the cluster then
traps stars from the underlying satellite before its disruption
causing the multiple populations with different kinematics. The models
by \citet{fel04} can reproduce the kinematics within $\omega$~Cen, but
did not model the populations in the object.

Here, we investigate a more general view of the fourth scenario. The
formation of a cluster complex takes about 10~Myr, which would also be
the time-scale for capturing field-stars.  This star-burst may have
been triggered by a previous perigalacticon passage of the dwarf
galaxy, but the final disruption of the host dwarf galaxy takes of the
order of 100~Myr or longer being the dynamical time scale near the
young Milky Way. 

While on the one hand, we see massive globular clusters with multiple
populations, on the other hand, we do find dwarf galaxies with
off-centre nuclei \citep{bin00}.  We want to investigate if these
off-centre nuclei, if formed out of merged star clusters in a massive
star-cluster complex inside a dwarf disc galaxy, can trap enough stars
from the underlying disc such that a measurable contamination by the
captured field stars with a different age and/or metallicity in the
merger object may be established.  The resulting off-centre massive
cluster may then be stripped from its host galaxy more easily than a
nucleus (i.e.\ in one of the perigalactica before the final
destruction of the host dwarf), but the destruction of the dwarf
galaxy is not part of the present work in which we concentrate on the
possible capture process of field stars.

To examine our scenario, we place a massive stellar super-cluster at
different radii in a dwarf disc galaxy. Note that, with this choice of
these radii, our intention is to probe a parameter-space of different
environments rather than to claim that these radii are actually the
formation sites of the off-centre nuclei).  Still, an off-centre
super-cluster can later sink to the centre or at least to much lower
distances from the centre (i.e.\ a few $100$~pc as found by
\citet{bin00}) due to dynamical friction.  We follow the orbit of this
cluster complex for a short time, i.e. the time needed to form a
merger object and measure the number of disc stars which become bound
to this object.  A detailed description of the numerical setup will
follow in the next section, while in Sect.~\ref{sec:results} we report
our results and show that under physically plausible circumstances
these objects can trap as much as half of their initial mass in disc
stars. I.e.\ one third of the stars in the final object are captured
older disc stars and are not formed in the star burst which formed the
object.
  
\section{Numerical Setup}
\label{sec:setup}

\begin{figure}
  \centering
  \epsfxsize=8.0cm
  \epsfysize=10.0cm
  \epsffile{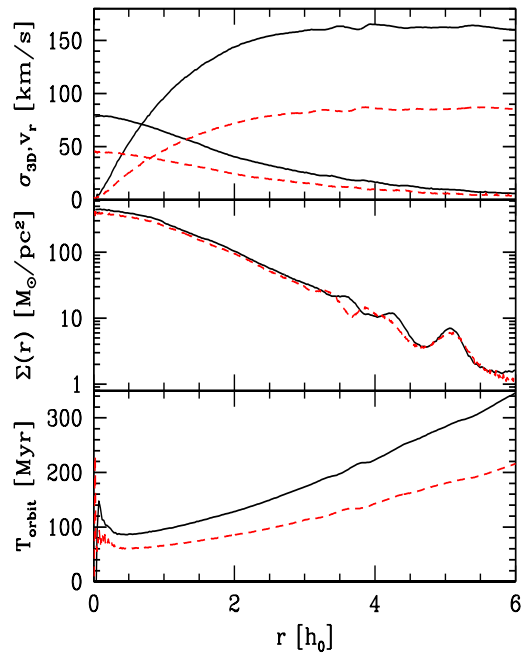}
  \caption{ Properties of the two disc models.  High mass disc is
    shown as solid lines (black on-line); low mass disc is shown as
    dashed lines (red on-line).  Upper panel: 3D-velocity dispersion
    (decaying curves) and rotational velocity (increasing curves).
    Middle panel: Surface density of the discs. Lower panel: Orbital
    time, i.e.\ time for a star to move once around the disc.}
  \label{fig:disc}
\end{figure}

We model a dwarf disc galaxy with mass $M_{\rm d}$, radial scale
length $h_{0}$ and vertical scale height $z_{0}$ as an exponential
disc in the $R$-direction and as isothermal sheets in the $z$-direction
\citep{spi42}:
\begin{eqnarray}
  \label{eq:disc}
  \rho(R,z) & = & \rho_{0}\ \exp(-R/h_{0})\ {\rm sech}^{2}(z/z_{0})
\end{eqnarray}
without a bulge.  The disc is modelled 'live' using 5,000,000
particles.  The disc is dynamically stable with a Toomre
$Q$-parameter of $1.5$ or higher.  The surrounding dark-matter halo is
modelled as an analytical logarithmic potential of the form
\begin{eqnarray}
  \label{eq:halo}
  \Phi(r) & = & v_{0}^{2}\ \ln(r^{2}+R_{\rm g}^{2}).
\end{eqnarray} 
The newly formed super-cluster is also modelled analytically as a
Plummer sphere orbiting in the disc at a radius $D$, with a Plummer
radius $R_{\rm pl}=25$~pc and a total mass of $M_{\rm c}$.

To investigate a range of parameters, we use three different models for
the disc and two different masses for the cluster complex.  In each
combination of models, the distance to the centre is varied as $D=0.5$,
$1.0$, $2.0$, $3.0$ and $5.0$ scale-lengths ($h_{0}$) of the disc.

As models for the disc galaxies, we use parameters suggested by the
observations of bulge-less dwarf disc-galaxies. For our low-mass
galaxy we adopt parameters from NGC~3274 taken from \citet{blo02},
\citet{swa02a} and \citet{swa02b} -- namely $M_{\rm d} =
10^{9}$~M$_{\odot}$, $h_{0} = 0.5$~kpc, $z_{0} = 100$~pc, $v_{0} =
80$~km\,s$^{-1}$ and $R_{\rm g} = 1$~kpc.  For our high-mass galaxy,
we used $M_{\rm d} = 10^{10}$~M$_{\odot}$, $h_{0} = 1.5$~kpc, $z_{0} =
250$~pc, $v_{0} = 150$~km\,s$^{-1}$ and $R_{\rm g} = 2.5$~kpc.  The
third disc model has the same parameters as the first one, but with
initially $10$~\% of the stars located in a thin, kinematically cold
sub-disc with the same scale length but a maximum height of $100$~pc.
The radial and azimuthal velocity dispersion are artificially reduced
by a factor $25$ and the vertical dispersion is adjusted to the
maximum height of $100$~pc.  The existence of such a cold population
would be expected given that a substantial number of stars are born in
small clusters with internal velocity dispersions $\simless 10$~km/s
\citep{kro02}.  {The velocity dispersion, rotational velocity,
surface density as well as the orbital timescale of the two disc
models are shown in Fig.~\ref{fig:disc}.} 

The final mass of the super-cluster is either $10^{6}$~M$_{\odot}$ or
$10^{7}$~M$_{\odot}$, and is linearly increased over a crossing-time
of the super-cluster from zero to the final value at the beginning of
the simulations.  The crossing-times of the super-clusters are $11.7$
and $3.7$~Myr, respectively.  { The linear mass increase is chosen
because it is the lowest-order approximation of a very complex process
during which the gas from which the cluster-complex forms is brought
together rapidly through large-scale supersonic motions. A
time-resolved increase in the super-cluster potential is required as
stars can only be trapped during its formation process.  At later
stages stars which enter the potential of the super-cluster will gain
as much energy as needed to leave the cluster on the other side again.
Effects like tidal capture need an interaction between the disc stars
and the stars of the super-cluster and are not treated by our simple
analytical treatment of the cluster potential, and are in any case
extremely rare.  On the other hand, it would also be unphysical to
introduce the super-cluster potential instantaneously.  We comment on
the results from a different representation of the formation process
at the end of the next section. The results are virtually
indistinguishable, showing that for the trapping process it is the
time-scale of the potential growth which is important, rather than the
functional form of potential growth.}

The simulations are carried out with the particle-mesh code Superbox
\citep{fel00}, the set-up of the grids being explained in the next
section.

\section{Results}
\label{sec:results}

\begin{figure}
  \centering
  \epsfxsize=8.0cm
  \epsfysize=8.0cm
  \epsffile{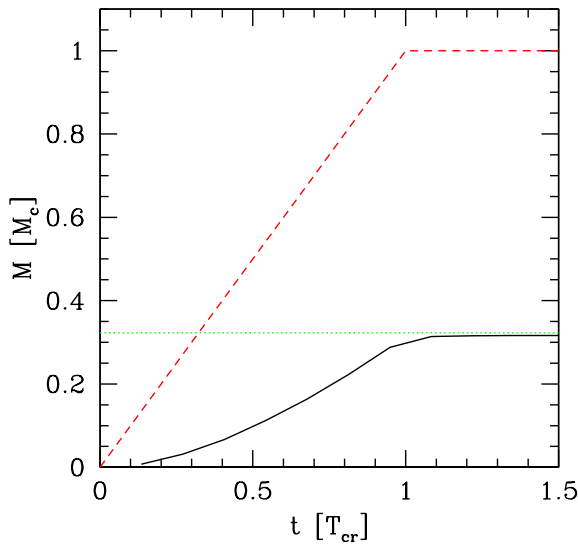}
  \caption{ This figure shows the growth in mass of the
    super-cluster compared with the growth in mass of the trapped
    stars for the model with $M_{\rm d} = 10^{9}$~M$_{\odot}$, $M_{\rm
      c} = 10^{7}$~M$_{\odot}$ and $D = 1 h_{0}$.  Mass is measured in
    units of $M_{\rm c}$ and time in units of the crossing time
    ($T_{\rm cr}$) of the super-cluster.  Solid line (black on-line)
    shows the trapped stars, dashed line (red on-line) the mass of the
    super-cluster potential and dotted line (green on-line) the mean
    trapped mass determined from all data-points until $200$~Myr.}
  \label{fig:grow}
\end{figure}

\begin{figure*}
  \centering
  \epsfxsize=8.0cm
  \epsfysize=8.0cm
  \epsffile{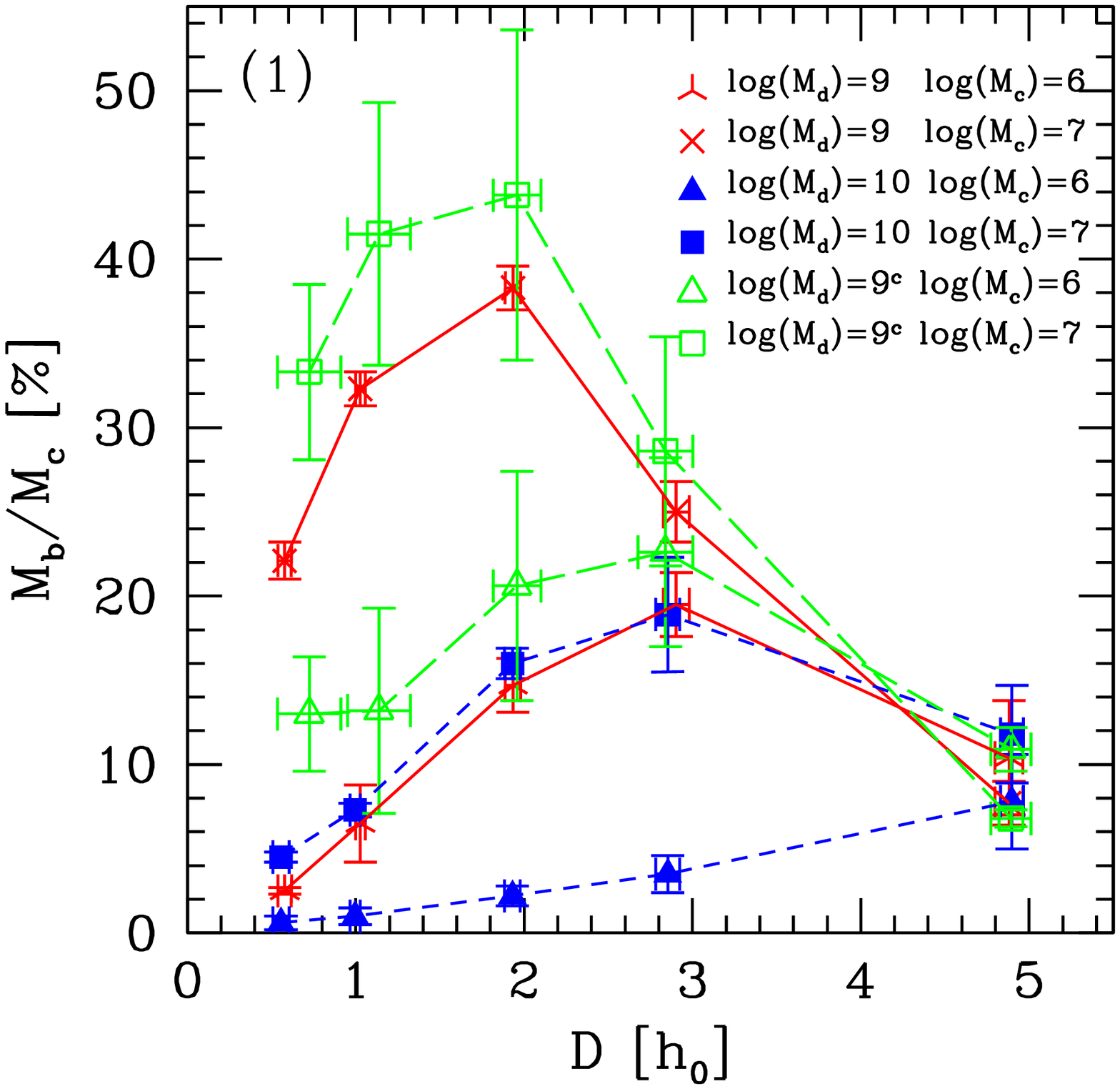}
  \epsfxsize=8.0cm
  \epsfysize=8.0cm
  \epsffile{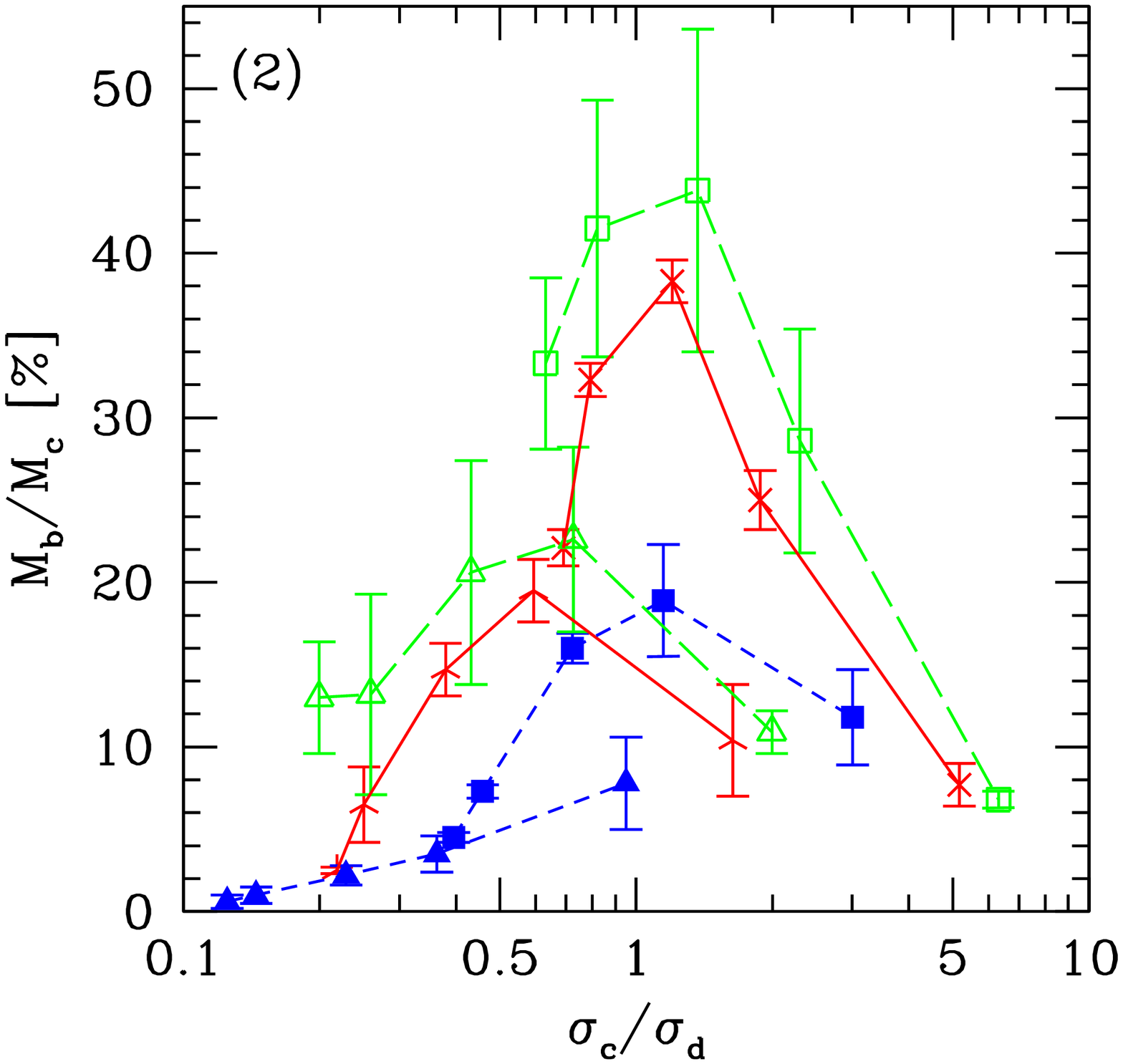}
  \epsfxsize=8.0cm
  \epsfysize=8.0cm
  \epsffile{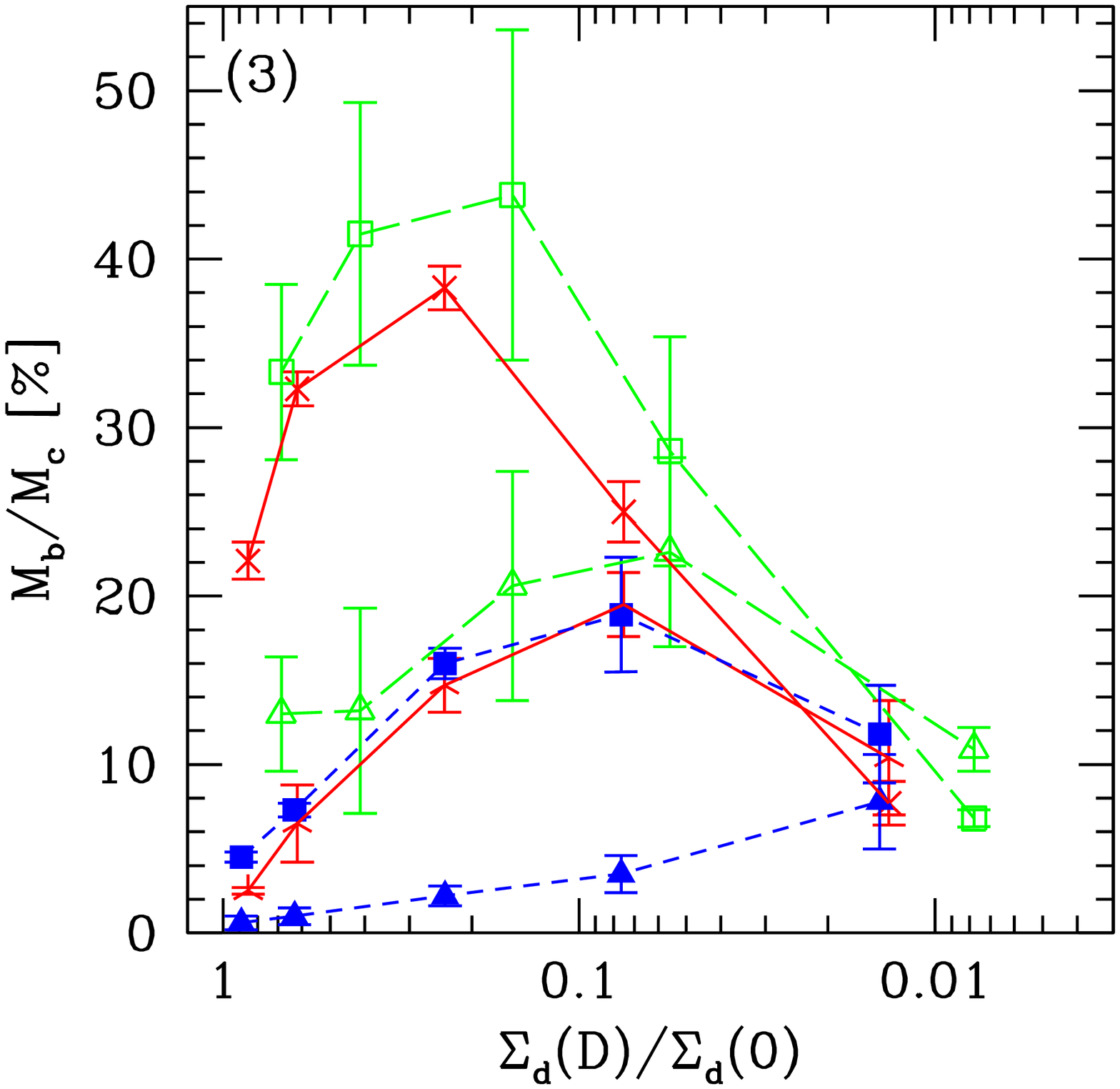}
  \epsfxsize=8.0cm
  \epsfysize=8.0cm
  \epsffile{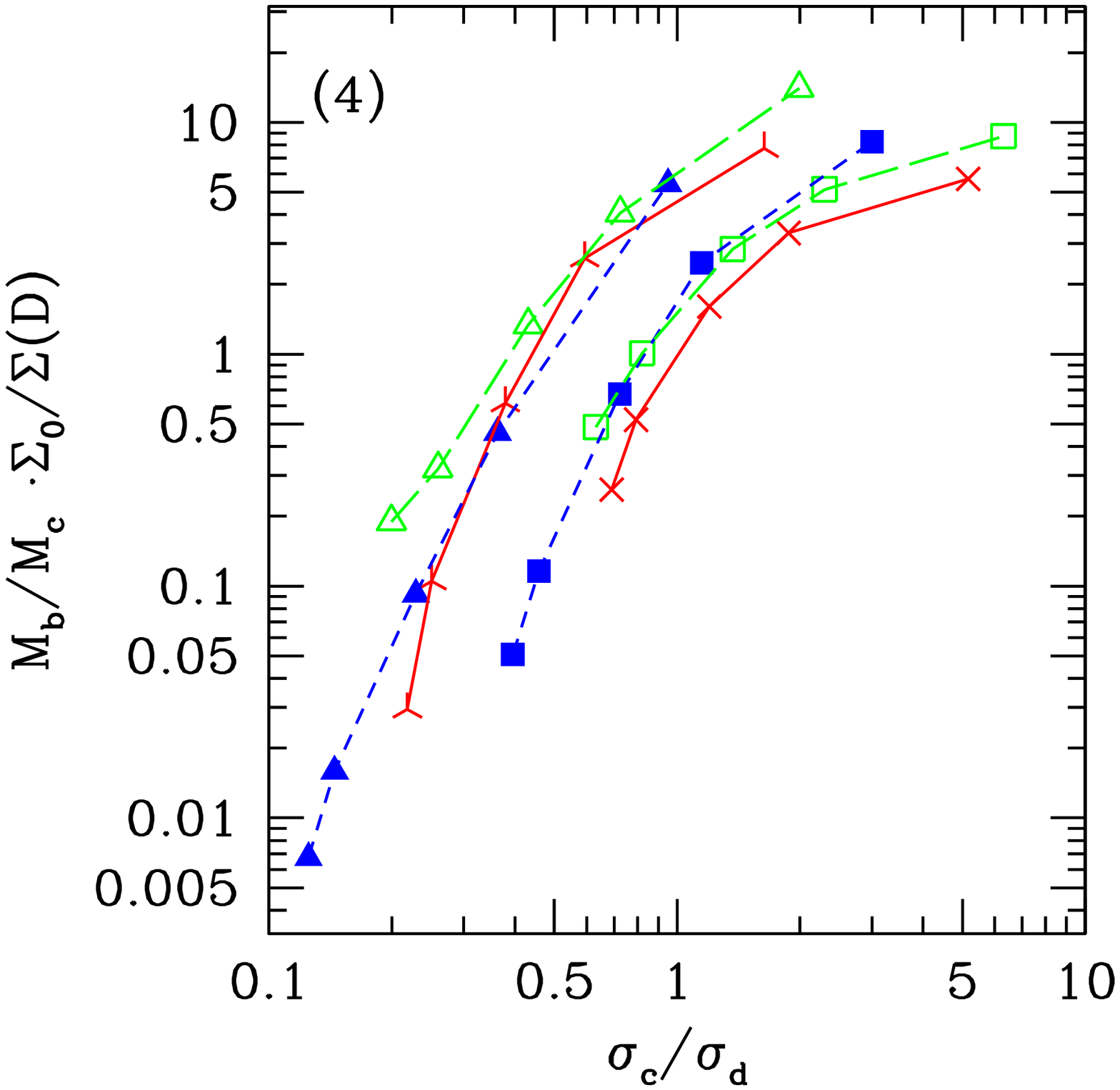}
  \caption{First panel: Ratio between the mass of the trapped disc
    stars $M_{\rm b}$ and the initial mass of the super-cluster
    $M_{\rm c}$ plotted against the radial distance $D$ to the disc
    centre in units of the radial scale length of the disc ($h_{0}$).
    Tri-pointed stars ($M_{\rm c} = 10^{6}$~M$_{\odot}$) and crosses
    ($M_{\rm c} = 10^{7}$~M$_{\odot}$) are the results derived in the
    low mass disc case ($M_{\rm d} = 10^{9}$~M$_{\odot}$, connected
    with solid lines, red on-line). Filled triangles ($M_{\rm c} =
    10^{6}$~M$_{\odot}$) and squares ($M_{\rm c} =
    10^{7}$~M$_{\odot}$) are the results of the high mass disc
    simulations ($M_{\rm d} = 10^{10}$~M$_{\odot}$, connected with
    short dashed lines, blue on-line).  Finally, open triangles
    ($M_{\rm c} = 10^{6}$~M$_{\odot}$) and open squares ($M_{\rm c} =
    10^{7}$~M$_{\odot}$) denote the results of the low mass disc
    simulations with $10$\,\% of kinematically cold stars (connected
    with long dashed lines, green on-line).  Second panel shows the
    same results now plotted against the ratio of the internal
    velocity dispersion of the cluster complex to the local
    dispersion of the disc stars.  It is clearly visible that the
    number of trapped stars increases with decreasing velocity
    dispersion of the disc stars until a ratio of about $1$ is
    reached.  Then the number of trapped stars decreases again because
    now the surface density of the disc stars drops significantly.
    This is shown in the third panel where the results are plotted
    against the surface density of the disc measured in units of the
    central surface density.  Finally, in the fourth panel, we multiply
    our results by the ratio of the surface densities.  { This panel
    shows that the turn-over in the previous panels is due to the
    strong decrease in surface density and may hint that if one has a
    high density and low velocity dispersion environment, a
    super-cluster could trap more than its initial mass (see
    discussion in the main text).}}
  \label{fig:results}
\end{figure*}

\begin{figure*}
  \centering
  \epsfxsize=8.0cm
  \epsfysize=8.0cm
  \epsffile{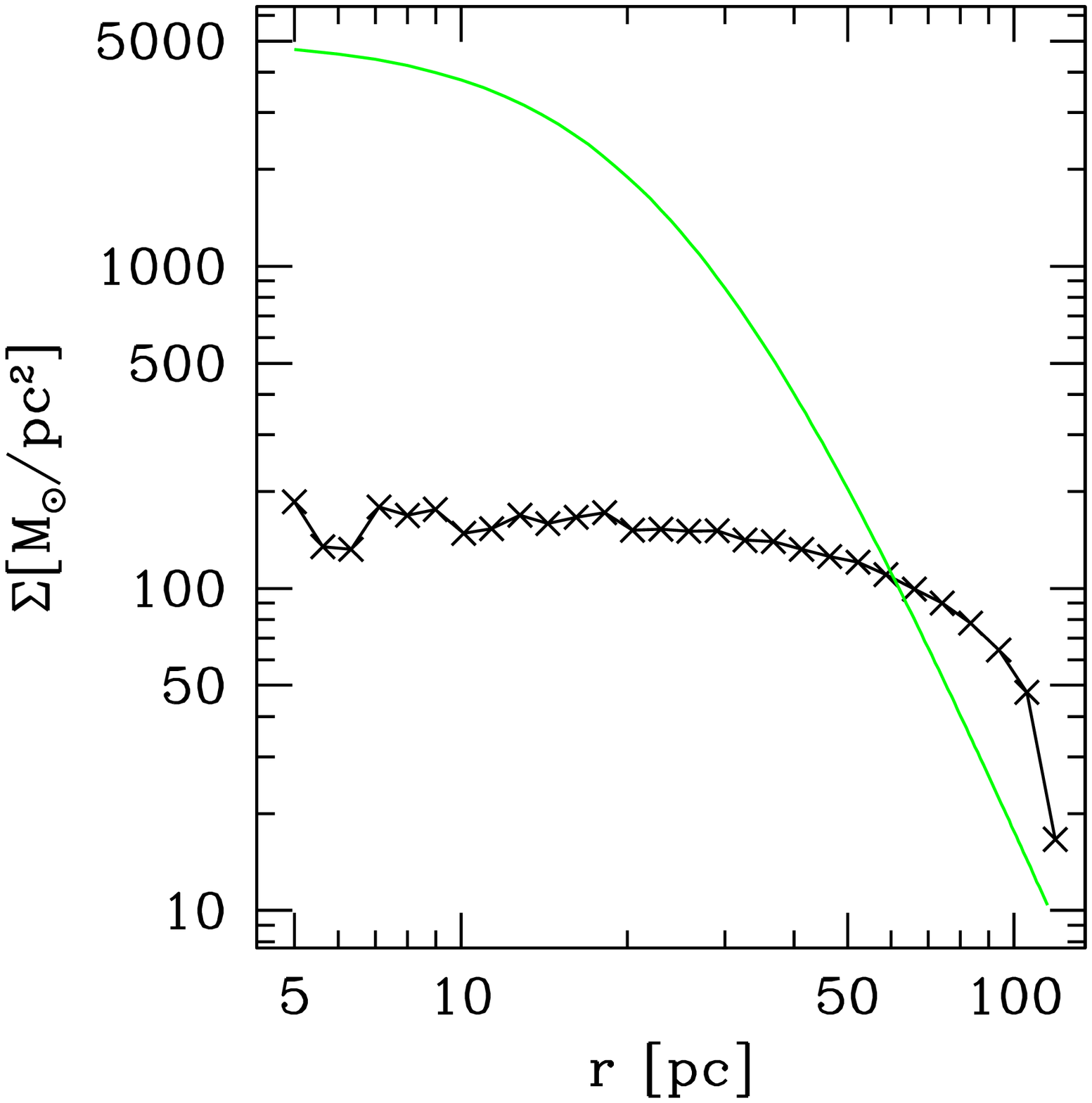}
  \epsfxsize=8.0cm
  \epsfysize=8.0cm
  \epsffile{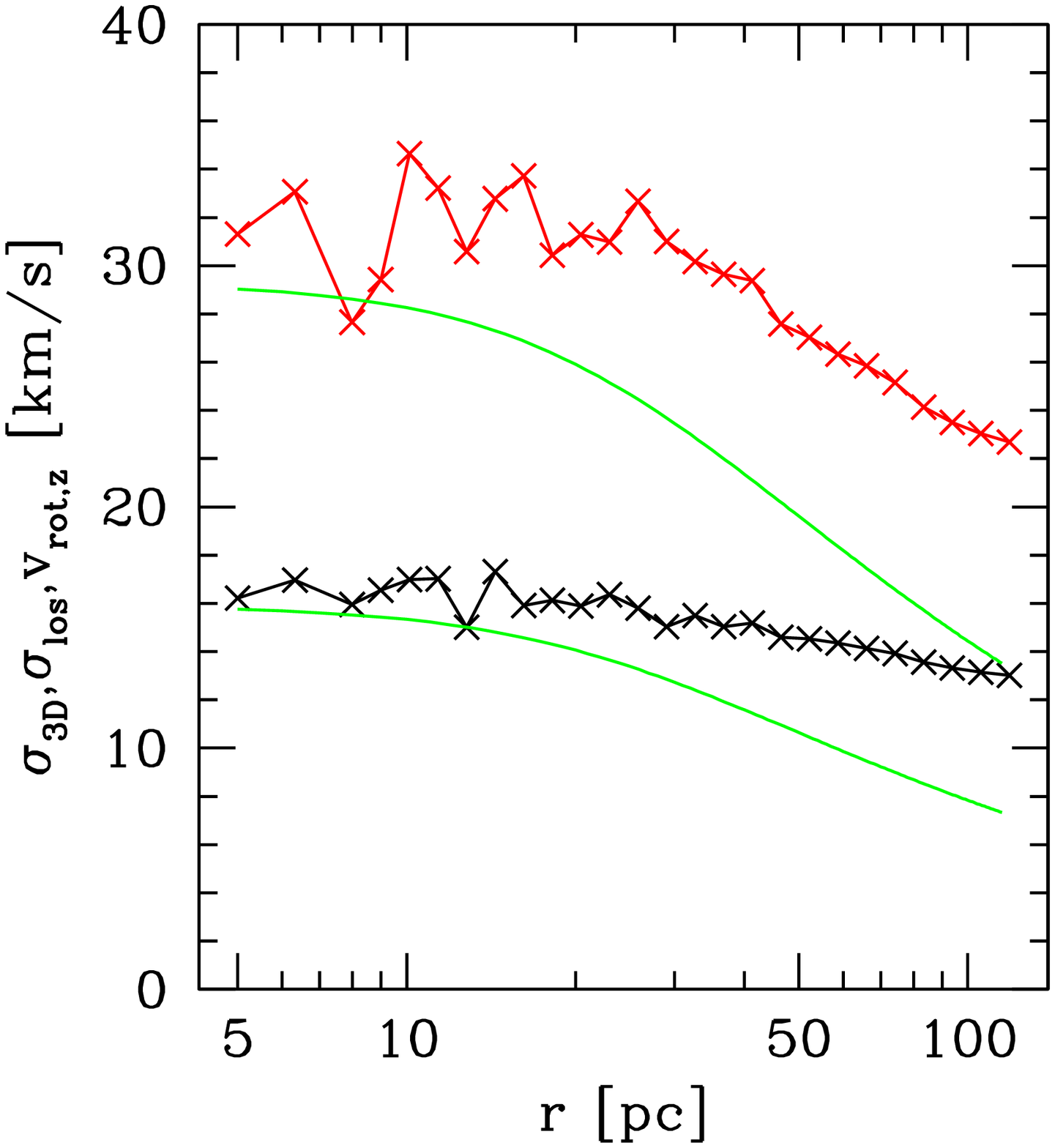}
  \caption{ { Distribution of the trapped stars for the low-mass
      disc, massive super-cluster and one scale-length distance
      simulation.}  Left panel: Surface
    density of the trapped stars. Connected crosses show the surface
    density of the trapped stars, while the solid line (green on-line)
    shows the surface density profile the pure analytical
    super-cluster would have (Plummer sphere).  Right panel: Upper
    curves (red and green on-line): 3D-velocity dispersion; lower
    curves (black and green on-line): line-of-sight velocity
    dispersion. Connected crosses show the data of the trapped stars
    while solid lines are the profiles of the analytical
    super-cluster.}
  \label{fig:profile}
\end{figure*}

The super-cluster is placed at different radii, $D$, on a circular
orbit with the circular speed at that radius.  The mass of the
super-cluster is increased linearly, as described above.  After that,
the mass of the super-cluster stays constant for the rest of the
simulation.  We run every model for $200$~Myr and determine the number
of trapped particles every $10$~Myr.  { So even our first time
slice is already after the formation of the super-cluster took place
or at the very end of it.  This means the trapping process is already
over when we investigate the particle distribution for the first
time.}  A particle is flagged as trapped if it has negative energy
with respect to the super-cluster potential and is located within the
tidal radius of the super-cluster.  We report the average number of
particles bound to the cluster determined in all $20$ time-slices
investigated.  This should ensure that the mass trapped at the
formation stays approximately constant and is not just composed of
'bound' transients. Still, most disc stars are just loosely bound.
This fact, as well as the choice of an analytical potential for the
super-cluster, the fixed time-step ($0.05$~Myr) and the fixed
resolution of the Superbox-grids (small disc: $17$~pc within
$0.5$~kpc, $67$~pc within $2$~kpc and $267$~pc beyond that; large
disc: $50$~pc within $1.5$~kpc, $200$~pc within $6$~kpc and $800$~pc
beyond that), gives the number of bound stars in a statistical sense.
We choose the $200$~Myr simulation time as this is long compared to
the crossing-time of the super-cluster and the time-scale on which the
disc stars get trapped.  This ensures that the effect of trapping
stars is not a transient feature.  It also spans about or more than
one orbital period of the super-cluster around the disc and is the
approximate time-scale between a first flyby and the final merging of
a disc galaxy with another disc galaxy.  But it is short compared to
the time-scale the newly formed object would need to sink to the
centre of the disc and form a nucleus or bulge, a process we do not
study with this contribution.

{ The detailed trapping process during the formation of the
super-cluster is not the subject of this study which concentrates on
quantifying the general outcome for the first time.  Basically, for a
given effective potential, stars at some particular location within
the final tidal radius of the super-cluster are trapped if they have a
velocity too low to carry them out of the final potential of the
super-cluster, and if the time-scale for the change of the potential is
shorter than or comparable to their crossing time through the
super-cluster.  Fig.~\ref{fig:grow} plots the mass of trapped stars as
a function of time (measured in crossing times of the super-cluster)
in the initial period during which the super-cluster forms for the low
mass disc ($M_{\rm d} = 10^{9}$~M$_{\odot}$), massive super-cluster
($M_{\rm c} = 10^{7}$~M$_{\odot}$) and $D = 1 h_{0}$ case.  The
trapped mass increases rapidly during the formation time of the
super-cluster.  After the super-cluster has reached its final mass,
essentially no late-comers are trapped.  To understand the evolution
of orbits of stars lying further away than the tidal radius would mean
a more in-depth analysis which will be the subject of a later
contribution.}

In Figure~\ref{fig:results} we show the ratio between the mass of the
trapped disc stars $M_{\rm b}$ and the mass of the super-cluster.  One
clearly sees that the number of trapped disc stars is not negligible.
The peak values in all parameter sets are always above $10$~per cent
of the super-cluster mass $M_{\rm c}$.  In the model where we added a
dynamically cold population, the peak value even reaches approximately
$45$~per cent.  In other words, in the final merger object, a very
massive star cluster, one out of three stars was not formed in the
star-burst which created the object.  While the original stars of the
super-cluster should all have the same metallicity and age, the
trapped stars should show a mix of the populations, metallicities and
ages of the underlying disc stars.  Most of these stars should stay
within the object even if the surrounding disc is destroyed by
interactions.  { The small error-bars in our results indicate that
except for some statistical fluctuations of very loosely bound stars
our results are not a transient feature as long as the disc potential
stays smooth.  In the simulations with an initially cold population, the
error-bars are larger because the natural heating-up of the cold stars
leads to density fluctuations like bars and spiral arms with density
contrasts of the order of or even higher than our object, acting as
disruptive forces on our super-cluster.}

Figure~\ref{fig:results} also shows that the peak value arises
approximately when the velocity dispersion of the disc stars equals
the velocity dispersion of the super-cluster.  The number of trapped
stars then decreases again because the surface density of the disc
drops.  With adding $10$~per cent of kinematically cold stars, i.e.\
the velocity dispersion of these stars is reduced by a factor of $25$,
one is able to increase the number of trapped stars by keeping the
same surface density (open symbols in Figure~\ref{fig:results}).  But
such a cold population is not a stable configuration.  Within a few
galactic orbits (a few~$\times 0.1 - 1$~Gyr) the stars of the cold
sub-population reach the same velocity dispersion as the other disc
stars.  In the much-more massive Milky Way this is evident in the
age--velocity dispersion relation (Kroupa 2002).  This time-scale is
long compared with the trapping time which is of the order of the
formation time of the super-cluster.  I.e.\ the cold sub-population
gets trapped preferentially and it is possible for it to show an
age-spread of about one~Gyr when compared to the endemic
population.

In the last panel of Figure~\ref{fig:results}, we show that the
turn-over in the ratio of trapped stars is really due to the
decreasing surface density.  When we divide our results by the ratio
between the actual surface density and the central one, the
turn-over disappears, showing a mild flattening towards an increasing
velocity-dispersion ratio instead (compare panels 2 and 4).  It also
shows that if the velocity dispersion of the surrounding stars is much
lower than the internal dispersion of the super-cluster for a constant
surface density, the super-cluster could trap more stars than its
initial mass.  This may point towards a possible model for $\omega$~Cen
which has, if the present notion is correct, $M_{\rm b}/M_{\rm c} =
70\%/25\% = 2.8$ \citep{hil00}; the most metal-rich 5~per cent of its
population may then have originated according to the third scenario
(accreted gas, Sec.~\ref{sec:intro}). 

{ One may argue that with our setup we create the mass of the
super-cluster out of nowhere.  This is true but since the mass ratio
between the forming cluster and the disc galaxy is small, the
adjustment of the host potential through the mass redistribution as a
result of cluster formation is a tiny effect.  Also the linear
increase of the mass of the super-cluster is only the lowest-order
approximation.  We therefore took one of our setups as described above
(small disc, heavy super cluster, one scale-length distance)  and
changed the way the super-cluster potential grows: the mass is taken
to be constant from the beginning of the simulation (mass
conservation), but the scale-length (Plummer radius) of the
super-cluster is shrunk from a disc scale length (i.e.\ the mass is
distributed over the whole disc extension) to its final value 
exponentially (slow at the beginning and very fast at the end) in the
following way:
\begin{eqnarray}
  \label{eq:expo}
  R_{\rm pl}(t) = (R_{\rm ini} - R_{\rm final}) \cdot \left( 1 -
    \exp(t-T_{\rm cr}) \right) + R_{\rm final}.
\end{eqnarray}
$R_{\rm pl}$ is the Plummer radius of the growing potential, $R_{\rm
  ini}$ and $R_{\rm final}$ are its initial ($1 h_{0}$) and final
($25$~pc) value.  $T_{\rm cr}$ denotes the crossing time of the
super-cluster.  This process models the assembly of a molecular cloud
from the interstellar medium spread throughout a large fraction of the
galaxy.}

Comparing the results of these two different ways to build-up the
super cluster potential shows no difference:  While in the first
variant we trap $33.2 \pm 1.0$~\% of the initial mass in disc stars,
the new build-up version gives us $33.3 \pm 1.0$~\%.

{ We also tested the robustness of our results with linear growth
of the potential and different time-scales ($1$--$20$~Myr) for the
potential to reach its final value and found no significant change of
the results.} 

Finally, we plot the radial profiles of the trapped stars in one of
our simulations in Figure~\ref{fig:profile}.  The surface density
distribution of the trapped stars has a large effective (half-light)
radius of about 90~pc.  It is much flatter and lower than the
distribution of the initial super-cluster.  The velocity dispersion of
the trapped stars has the same central value as the super-cluster but
decreases much more slowly, the trapped stars thereby forming a
dynamically hotter population in the outer parts of the final object.

\section{Conclusions}
\label{sec:conclus}

With our suite of simulation, we have shown that it is possible that
very massive newborn star clusters or star cluster complexes, which
later merge into one massive object, trap a detectable amount of
underlying stars of the host galaxy in which they are born.
With our choice of parameters, a super-cluster may capture up to one
foreign star for every two home stars ($M_{\rm b}/M_{\rm c}\approx
0.5$).

We also showed that, in dwarf disc galaxies with exponential profiles,
this effect is largest when the velocity dispersions of the
super-cluster and the disc stars are comparable.  This maximum arises
due to the fact that in the outer parts of the disc galaxy, where the
velocity dispersion is sufficiently low, the surface density is also
very low, i.e.\ the super-cluster simply does not find enough stars to
trap.  This is shown in the last panel of Fig.~\ref{fig:results}.  One
may also read this panel such that, for a low field velocity
dispersion and a constant surface density the super-cluster traps
several times its own mass.

The question arises if there are environments with low velocity
dispersions (i.e.\ dynamically cold) which have high densities so that
the newborn cluster could even trap several times its own initial mass
$M_{\rm b}/M_{\rm c}>2$.  { These environments are probably not
stable dynamically and would heat-up within a few orbital times of the
disc, but the time-scale of trapping is much shorter.  It is of the
order of the formation time of the newborn super-cluster, namely a
few~Myr.  Thus, even in this case the original and captured
populations may show substantial age and metallicity differences.
}

We also expect this effect for massive star clusters born in dSph
systems.  Even though these systems are dynamically hot given their
visible mass, they have velocity dispersions of only about
$10$~km\,s$^{-1}$.  If a very massive super-cluster formed in such an
object, it would trap a significant amount of old stars.  Another
dwarf galaxy nearby is the Large Magellanic Cloud (LMC).  With a
velocity dispersion of about $30$~km\,s$^{-1}$, we expect that a very
massive cluster (say $10^{6}$~M$_{\odot}$) traps a few per cent of its
initial mass from the LMC.  With better observational data of single
stars in the LMC clusters, this could be a measurable effect.

As a final remark, we point out that the surface density distribution
of the trapped stars is much flatter than the stars of the initial
object and these stars form a dynamically hotter sub-system in the
outer parts of the object. 

\vspace{6mm}

\noindent {\bf Acknowledgements:}
MF gratefully acknowledges financial support through DFG-grant
KR1635/5-1 and PPARC.

\label{lastpage}

\end{document}